\documentclass[11pt]{article}
\usepackage{epsfig} 
\newcommand{\third}{\mbox{\small{$\frac{1}{3}$}}} 
 
\newcommand{\pitwo}{\mbox{\small{$\frac{\pi}{2}$}}} 
\newcommand{\pisix}{\mbox{\small{$\frac{\pi}{6}$}}} 
\newcommand{\MSbar}{\overline{\mbox{MS}}} 
\newcommand{\MSbars}{\overline{\mbox{\footnotesize{MS}}}} 
\newcommand{\MOMgggs}{\mbox{\footnotesize{MOMggg}}}
\newcommand{\wMOMggs}
{{\widetilde{\mbox{\footnotesize{MOM}}}}\mbox{\footnotesize{gg}}}
\newcommand{\MOMhs}{\mbox{\footnotesize{MOMh}}}
\newcommand{\MOMqs}{\mbox{\footnotesize{MOMq}}}
\newcommand{\MOMis}{\mbox{\footnotesize{MOMi}}}
\newcommand{\Nf}{N_{\!f}}

\begin{document}
\title{Three loop QCD MOM $\beta$-functions}
\author{J.A. Gracey, \\ Theoretical Physics Division, \\ 
Department of Mathematical Sciences, \\ University of Liverpool, \\ P.O. Box 
147, \\ Liverpool, \\ L69 3BX, \\ United Kingdom.} 
\date{} 
\maketitle 

\vspace{5cm} 
\noindent 
{\bf Abstract.} We present the full expressions for the QCD $\beta$-function
in the MOMggg, MOMq and MOMh renormalization schemes at three loops for an
arbitrary colour group in the Landau gauge. The results for all three schemes
are in very good agreement with the $SU(3)$ numerical estimates provided by
Chetyrkin and Seidensticker.

\vspace{-15.2cm}
\hspace{13.5cm}
{\bf LTH 911}

\newpage

Renormalization schemes can be divided into various classes. For instance, they
can be split into physical and unphysical or mass dependent and mass 
independent schemes. One of the main schemes which is used in perturbative 
quantum field theories is the minimal subtraction scheme, \cite{1}, or its more
widely used extension which is the modified minimal subtraction scheme and
abbreviated by $\MSbar$, \cite{2}. A reason the latter is predominant is that 
it is the scheme in which one can compute to very high orders in perturbation 
theory. For example, the four loop QCD $\beta$-function, \cite{3,4}, represents
the current state of the art for the $\beta$-function of the theory of strong 
interactions. Indeed computations proceed to this high order because $\MSbar$ 
is a mass independent renormalization scheme and thus one merely needs to be 
able to compute massless Feynman diagrams. This is possible even for vertex 
functions where an external momentum can be nullified to allow the simple 
extraction of the poles in the regulator. One consequence of this is that 
techniques such as integration by parts in $d$ spacetime dimensions, 
\cite{5,6,7}, which are implemented within computer algebra programmes such as 
{\sc Form}, \cite{8}, can be used systematically. However, a drawback of the 
$\MSbar$ scheme is that the subtraction point for the vertex renormalization, 
and hence the definition of the coupling constant renormalization constant, is 
at an unphysical point of the vertex Green's function. Therefore, despite its 
attractiveness with regard to computability $\MSbar$ is an unphysical scheme 
and thus one could encounter infrared inconsistencies if one were to compare 
with the nonperturbative structure of the vertex Green's function. By contrast 
it would be more appropriate to define the subtraction point of the Green's 
function at a physical point which avoids any potential infrared ambiguities
and thus is more useful for nonperturbative measurements such as those carried 
out using lattice regularization. Of necessity such schemes whilst being 
physical introduce a mass scale within Feynman diagrams even in a theory which 
was originally massless in contrast to $\MSbar$. Moreover, nullification of an
external momentum of a vertex is not permitted. As such these effectively 
massive Feynman integrals are more difficult to compute analytically, and thus
the renormalization group functions of QCD in such schemes are not available to
as high a loop order as the $\MSbar$ scheme.  

One class of physical and mass dependent schemes which is of importance to 
lattice gauge theory computations is the set referred to as MOM which denotes
the family of momentum subtraction schemes, \cite{9}. Though the syntax MOM 
indicates a specific momentum configuration of the QCD $3$-point vertices. This
is the situation where the squares of all the momenta coming in through the 
three vertices are equal to one mass squared value. This is known as the 
symmetric subtraction point. The vertex renormalization is performed in such a 
way that after the MOM renormalization constants are set then there are no 
$O(a)$ corrections at the subtraction point where $a$~$=$~$g^2/(16\pi^2)$ and 
$g$ is the coupling constant of gluon with itself or the quarks and 
Faddeev-Popov ghost fields. In other words if the integrals are dimensionally 
regularized, which we will use throughout, then not only are the poles in 
$\epsilon$ removed, where $d$~$=$~$4$~$-$~$2\epsilon$, but also the finite part
at the subtraction point. Indeed for the $2$-point functions of the theory the 
same criterion is used so that at their subtraction point there are no $O(a)$
contributions after the wave function and gauge parameter MOM renormalization 
constants are set. This outline is a summary of the original MOM scheme
definition of \cite{9}. There the full one loop renormalization of QCD was 
carried out. However, as one has several $3$-point vertices in the QCD 
Lagrangian when there is a linear covariant gauge fixing, it was noted in 
\cite{9} that there are three types of MOM schemes. Each is associated with a 
particular $3$-point vertex and are denoted by MOMggg, MOMq and MOMh. These 
refer respectively to the schemes derived by ensuring triple gluon, quark-gluon
and ghost-gluon vertices are subtracted by the MOM criterion. Once each of 
these schemes is defined the structure of the remaining $n$-point functions of 
the theory are determined by the Slavnov-Taylor identities of QCD, \cite{9}. 
Indeed in \cite{9} the full one loop renormalization of QCD in each of the 
three MOM schemes was discussed. 

Beyond one loop one runs into the difficulty of calculability. This is
illustrated in particular in \cite{10}. There the two loop relation between the
$\MSbar$ coupling constant and that of each of the three MOM schemes was 
derived {\em numerically}. This was partly because the basic scalar master two 
loop Feynman integrals for the symmetric subtraction point for the $3$-point 
topologies were not fully available. A subset of the necessary integrals have
been evaluated in \cite{11,12,13} but the final master was only constructed as 
a corollary of the results of \cite{14}. There the basic $3$-point functions 
were determined for the more general configuration of all three squared 
external momenta being independent. Therefore in the absence of these results 
at the time, the approach of \cite{10} was to approximate the $3$-point 
integrals by using a large mass expansion with respect to one of the 
independent momentum squared. This allowed for the use of the {\sc Mincer}
algorithm, \cite{15}, and application of the {\sc Exp} package. With a 
sufficient number of terms in the expansion in the appropriate parameter an 
approximation was determined for the $3$-point vertices to two loops. Equipped 
with this the mapping of the coupling constants between MOM and $\MSbar$ 
schemes was established. Hence through the renormalization group equation the 
{\em three} loop MOM scheme $\beta$-functions were given numerically for 
$SU(3)$, \cite{10}. This was possible as the three loop $\MSbar$ 
$\beta$-function was already available, \cite{16,17,18,19,20,21}. Whilst this 
represents the current status of these $\beta$-functions in the MOM schemes it 
is the purpose of this article to report on the exact evaluation of the three 
loop MOM $\beta$-functions for an arbitrary colour group. This is possible with
the advance not only in computer power but also because of the development of 
the Laporta algorithm, \cite{22}. This allows one to systematically write all 
Feynman integrals of a certain topology in terms of the basic scalar master 
integrals. As the two loop masters are available for the $3$-point functions, 
\cite{11,12,13,14}, it is merely a straightforward exercise to repeat the 
approach of \cite{10} and compute the vertex functions to two loops at the 
symmetric subtraction point. Although the difference here is to avoid any 
numerical approximation to the integrals, we will still follow the ethos of 
constructing the mapping between the parameters of both schemes. This will
allow us to use the same conversion technique to establish the precise 
structure of the three loop terms of the $\beta$-function in each of the three 
MOM schemes. We will focus specifically on the $\beta$-functions here as the 
limited space excludes us from reporting on the full structure of each of the 
three vertices at the symmetric subtraction point. However, this together with
the parameter mappings and the explicit forms of all the anomalous dimensions 
of QCD in each of the three separate schemes for an arbitrary linear covariant 
gauge and colour group will appear in a longer article, \cite{23}. Finally, as 
background we mention that the triple gluon vertex has been examined at two 
loops previously in the on-shell configuration in \cite{24}.

More specifically we have computed the three Green's functions which relate to 
each of the vertices associated with the renormalization schemes. These are
each written in terms of a set of Lorentz basis tensors built out of the two  
independent external momenta, $p$ and $q$, together with the metric tensor
$\eta_{\mu\nu}$ as well as $\gamma$-matrices in the case of the quark-gluon
vertex. Denoting the basis tensors for the $i$th vertex by 
${\cal P}^i_{(k) \, \{\mu_i\} }(p,q)$ where $\{\mu_i\}$ indicates the 
associated Lorentz indices for that vertex and $k$ labels the tensor, then we 
have 
\begin{eqnarray}
\left. \left\langle A^a_\mu(p) A^b_\nu(q) A^c_\sigma(-p-q)
\right\rangle \right|_{p^2 = q^2 = - \mu^2} &=& f^{abc} \sum_{k=1}^{14}
{\cal P}^{\mbox{\footnotesize{ggg}}}_{(k) \, \mu \nu \sigma }(p,q) \,
\Sigma^{\mbox{\footnotesize{ggg}}}_{(k)}(p,q) \nonumber \\
\left. \left\langle \psi^i(p) \bar{\psi}^j(q) A^c_\sigma(-p-q)
\right\rangle \right|_{p^2 = q^2 = - \mu^2} &=& T^c_{ij} \sum_{k=1}^{6}
{\cal P}^{\mbox{\footnotesize{qqg}}}_{(k) \, \sigma }(p,q) \,
\Sigma^{\mbox{\footnotesize{qqg}}}_{(k)}(p,q) \nonumber \\
\left. \left\langle c^a(p) \bar{c}^b(q) A^c_\sigma(-p-q)
\right\rangle \right|_{p^2 = q^2 = - \mu^2} &=& f^{abc} \sum_{k=1}^{2}
{\cal P}^{\mbox{\footnotesize{ccg}}}_{(k) \, \sigma }(p,q) \,
\Sigma^{\mbox{\footnotesize{ccg}}}_{(k)}(p,q) 
\label{vertdef}
\end{eqnarray}
where $\Sigma^i_{(k)}(p,q)$ is the scalar amplitude associated with each basis 
tensor at the symmetric point. As we are focusing on the structure of the 
vertices at the symmetric subtraction point then the momenta satisfy 
\begin{equation}
p^2 ~=~ q^2 ~=~ ( p + q )^2 ~=~ -~ \mu^2
\end{equation}
which imply
\begin{equation}
pq ~=~ \frac{1}{2} \mu^2 
\end{equation}
where $\mu$ is the mass scale introduced in dimensional regularization to
ensure the coupling constant remains massless in $d$ dimensions. It ought to be
noted that away from this point the tensor basis will be larger than that 
indicated in the decomposition. However, for the purpose of the construction of
the MOM $\beta$-functions the only tensor of interest, and hence its associated 
amplitude, is that which has the same Lorentz structure as the original vertex 
in the QCD Lagrangian. By renormalizability it is this channel which contains 
the poles in $\epsilon$ which have to be removed in the scheme of interest. 
Indeed the definition of the MOM type schemes is to render the $O(a)$ 
corrections absent after subtraction in the channel, or channels in the case of
the triple gluon vertex, with the divergences. The amplitudes of the remaining 
channels do not necessarily have no $O(a)$ contributions. As ultimately they do
not affect the $\beta$-functions but instead will be useful for lattice 
measurements of these Green's functions we will record them in a longer 
article, \cite{23}, where the full tensor basis will be provided.

To evaluate each of the amplitudes we use the method of projection where we
determine the linear combinations of the basis tensors which in turn 
determine the amplitude of interest. This combination is then applied to the
vertex and the resulting scalar integral is evaluated by using the Laporta
algorithm, \cite{22}. The major working tool in this respect was the use of the
{\sc Reduze} package, \cite{25}, which uses the symbolic manipulation formalism
of {\sc Ginac}, \cite{26}, which itself is written in C++. In order to handle 
the tedious algebra which arises with the rearrangement of the expressions 
resulting from each projection to a form in which the Laporta algorithm is
applied we used the symbolic manipulation language {\sc Form} \cite{8}. The
underlying Feynman diagrams were generated using the {\sc Qgraf} package,
\cite{27}, before the indices were appended and the Feynman rules substituted.
For each of the vertices in turn, triple gluon, quark-gluon and ghost-gluon, 
there were $8$, $2$ and $2$ one loop diagrams. At two loops there were 
respectively $106$, $33$ and $33$ graphs to determine. To perform the 
renormalization itself in the various schemes we followed the method of
\cite{21} where all the diagrams are determined in terms of bare parameters
which here are the coupling constant and gauge parameter of the linear gauge
fixing. Then the renormalized variables are introduced by rescaling with the
renormalization constants. The advantage of this approach is that one needs
only to generate the results for each of the Green's functions once and then
the renormalization constants for each scheme are deduced as simple corollaries
without having to regenerate the expressions for each individual diagram. 

Once each of the Green's functions of (\ref{vertdef}) has been renormalized in
their respective schemes then the $\beta$-function in each scheme is 
determined from, \cite{10},
\begin{equation}
\beta^{\mbox{$\MOMis$}} ( a_{\mbox{$\MOMis$}}, \alpha_{\mbox{$\MOMis$}} ) ~=~
\left[ \beta^{\mbox{$\MSbars$}}( a_{\mbox{$\MSbars$}} ) 
\frac{\partial a_{\mbox{$\MOMis$}}}{\partial a_{\mbox{$\MSbars$}}} \,+\,
\alpha_{\mbox{$\MSbars$}} \gamma^{\mbox{$\MSbars$}}_\alpha 
( a_{\mbox{$\MSbars$}}, \alpha_{\mbox{\footnotesize{$\MSbars$}}} ) 
\frac{\partial a_{\mbox{$\MOMis$}}}{\partial \alpha_{\mbox{$\MSbars$}}} 
\right]_{ \MSbars \rightarrow \MOMis }
\end{equation}
where $\mbox{i}$ represents ggg, q or h and $\gamma_\alpha(a,\alpha)$ is the
anomalous dimension of the linear covariant gauge parameter. The scheme the
renormalization group functions and variables are in are denoted by the labels.
Each of these $\beta$-functions requires the relation of the parameters $a$ and
$\alpha$ in each MOM scheme to the same parameters in the $\MSbar$ scheme. The 
identification mapping indicated on the right hand side is a reminder that the 
variables are initially in the $\MSbar$ scheme and must then be mapped back to 
their MOMi counterparts. To achieve this we define the relations between both
parameters by  
\begin{eqnarray}
a_{\MOMgggs}(\mu) &=& a_{\MSbars}(\mu) \left. \left[ \frac{\Pi_g^{\MOMgggs}(p)}
{\Pi_g^{\MSbars}(p)} \right]^3 \right|_{p^2 \, = \,-\, \mu^2}
\left[ \frac{\Sigma^{\mbox{\footnotesize{ggg}}}_{(1)\,\MSbars}(-\mu^2,-\mu^2)}
{\Sigma^{\mbox{\footnotesize{ggg}}}_{(1)\,\MOMgggs}(-\mu^2,-\mu^2)} \right]^2 
\nonumber \\
a_{\MOMqs}(\mu) &=& a_{\MSbars}(\mu) \left. \left[ \frac{\Pi_g^{\MOMqs}(p) 
\left( \Sigma^{\MOMqs}_q(p) \right)^2} {\Pi_g^{\MSbars}(p) \left( 
\Sigma^{\MSbars}_q(p) \right)^2} \right] \right|_{p^2 \, = \,-\, \mu^2}
\left[ \frac{\Sigma^{\mbox{\footnotesize{qqg}}}_{(1)\,\MSbars}(-\mu^2,-\mu^2)}
{\Sigma^{\mbox{\footnotesize{qqg}}}_{(1)\,\MOMqs}(-\mu^2,-\mu^2)} \right]^2 
\nonumber \\
a_{\MOMhs}(\mu) &=& a_{\MSbars}(\mu) \left. \left[ \frac{\Pi_g^{\MOMhs}(p) 
\left( \Sigma^{\MOMhs}_c(p) \right)^2} {\Pi_g^{\MSbars}(p) \left( 
\Sigma^{\MSbars}_c(p) \right)^2} \right] \right|_{p^2 \, = \,-\, \mu^2}
\left[ \frac{\Sigma^{\mbox{\footnotesize{gcc}}}_{(1)\,\MSbars}(-\mu^2,-\mu^2)}
{\Sigma^{\mbox{\footnotesize{gcc}}}_{(1)\,\MOMhs}(-\mu^2,-\mu^2)} \right]^2 
\label{ccdef}
\end{eqnarray}
where ggg, qqg and gcc indicate the respective triple gluon, quark-gluon and
ghost-gluon vertices and
\begin{equation}
\alpha_{\MOMis}(\mu) ~=~ \frac{Z^{\MOMis}}{Z^{\MSbars}}
\alpha_{\MSbars}(\mu) ~. 
\label{aldef}
\end{equation}
In the coupling constant definitions the contributions from the quark and
Faddeev-Popov ghost $2$-point functions are $\Sigma_q(p)$ and $\Sigma_c(p)$
respectively whilst $\Pi_g(p)$ is the contribution to the transverse part of 
the gluon $2$-point function. In (\ref{ccdef}) we have included the 
contribution from the MOMi scheme Green's functions in order that the overall 
normalization is consistent even though there are no $O(a)$ corrections in 
their expansion. Also in constructing the mapping between the parameters $a$ 
and $\alpha$ between schemes, (\ref{ccdef}) and (\ref{aldef}) are solved 
iteratively. 

As a consequence of this we can now record our main result which is the three 
loop expressions for the $\beta$-functions, $\beta^{\mbox{$\MOMis$}} 
(a,\alpha)$, in each of the three MOM schemes. Though due to space 
considerations we will only record the Landau gauge expressions. The full gauge
dependent results will be provided in \cite{23}. With the convention that the 
scheme given in the superscript on the left hand side is the scheme the 
variables are defined in via (\ref{ccdef}) and (\ref{aldef}), then we 
have
\begin{eqnarray}
\beta^{\mbox{$\MOMgggs$}}(a,0) &=& -~ \left[ \frac{11}{3} C_A - \frac{4}{3} 
T_F \Nf \right] a^2 ~-~ \left[ \frac{34}{3} C_A^2 - 4 C_F T_F \Nf 
- \frac{20}{3} C_A T_F \Nf \right] a^3 \nonumber \\
&& +~ \left[ \left[ 
209484 (\psi^\prime(\third))^2
- 279312 \pi^2 \psi^\prime(\third) 
+ 37087200 \psi^\prime(\third)
\right. \right. \nonumber \\
&& \left. \left. ~~~~~
+~ 368874 \psi^{\prime\prime\prime}(\third) 
+ 266359104 s_2(\pisix)
- 532718208 s_2(\pitwo)
\right. \right. \nonumber \\
&& \left. \left. ~~~~~
-~ 443931840 s_3(\pisix)
+ 355145472 s_3(\pitwo)
- 890560 \pi^4 
\right. \right. \nonumber \\
&& \left. \left. ~~~~~
-~ 24724800 \pi^2 
+ 416988 \Sigma 
- 30440124 \zeta(3) 
- 51650217
\right. \right. \nonumber \\
&& \left. \left. ~~~~~
+~ 1849716 \frac{\ln^2(3) \pi}{\sqrt{3}} 
- 22196592 \frac{\ln(3) \pi}{\sqrt{3}}
- 1986732 \frac{\pi^3}{\sqrt{3}}
\right] C_A^3 \right. \nonumber \\
&& ~~~~~+ \left[ \left[ 
1656000 \pi^2 \psi^\prime(\third) 
- 1242000 (\psi^\prime(\third))^2
- 38988864 \psi^\prime(\third)
\right. \right. \nonumber \\
&& \left. \left. ~~~~~~~~~~
-~ 134136 \psi^{\prime\prime\prime}(\third) 
- 220029696 s_2(\pisix)
+ 440059392 s_2(\pitwo)
\right. \right. \nonumber \\
&& \left. \left. ~~~~~~~~~~
+~ 366716160 s_3(\pisix)
- 293372928 s_3(\pitwo)
- 194304 \pi^4 
\right. \right. \nonumber \\
&& \left. \left. ~~~~~~~~~~
+~ 25992576 \pi^2 
- 8363088 \Sigma 
+ 43914960 \zeta(3) 
+ 49845132
\right. \right. \nonumber \\
&& \left. \left. ~~~~~~~~~~
-~ 1527984 \frac{\ln^2(3) \pi}{\sqrt{3}} 
+ 18335808 \frac{\ln(3) \pi}{\sqrt{3}}
+ 1641168 \frac{\pi^3}{\sqrt{3}}
\right] C_A^2 T_F \Nf \right. \nonumber \\
&& ~~~~~+ \left[ \left[ 
2045952 (\psi^\prime(\third))^2
- 2727936 \pi^2 \psi^\prime(\third) 
+ 11591424 \psi^\prime(\third)
\right. \right. \nonumber \\
&& \left. \left. ~~~~~~~~~~
+~ 44789760 s_2(\pisix)
- 89579520 s_2(\pitwo)
- 74649600 s_3(\pisix)
\right. \right. \nonumber \\
&& \left. \left. ~~~~~~~~~~
+~ 59719680 s_3(\pitwo)
+ 909312 \pi^4 
- 7727616 \pi^2 
\right. \right. \nonumber \\
&& \left. \left. ~~~~~~~~~~
+~ 2985984 \Sigma 
- 11943936 \zeta(3) 
- 8460288
\right. \right. \nonumber \\
&& \left. \left. ~~~~~~~~~~
+~ 311040 \frac{\ln^2(3) \pi}{\sqrt{3}} 
- 3732480 \frac{\ln(3) \pi}{\sqrt{3}}
- 334080 \frac{\pi^3}{\sqrt{3}}
\right] C_A T_F^2 \Nf^2 \right. \nonumber \\
&& ~~~~~+ \left[ \left[ 
786432 \pi^2 \psi^\prime(\third) 
- 589824 (\psi^\prime(\third))^2
- 442368 \psi^\prime(\third)
\right. \right. \nonumber \\
&& \left. \left. ~~~~~~~~~~
-~ 262144 \pi^4 
+ 294912 \pi^2 
- 82944
\right] T_F^3 \Nf^3 \right. \nonumber \\
&& ~~~~~+ \left[ \left[ 
4758912 \psi^\prime(\third)
- 456192 \psi^{\prime\prime\prime}(\third) 
+ 1216512 \pi^4 
- 3172608 \pi^2 
\right. \right. \nonumber \\
&& \left. \left. ~~~~~~~~~~
+~ 16422912 \Sigma 
- 24634368 \zeta(3) 
+ 23421312
\right] C_A C_F T_F \Nf \right. \nonumber \\
&& ~~~~~+ \left[ \left[ 
165888 \psi^{\prime\prime\prime}(\third) 
- 442368 \pi^4 
- 5971968 \Sigma 
+ 8957952 \zeta(3) 
\right. \right. \nonumber \\
&& \left. \left. ~~~~~~~~~~
-~ 7091712
\right] C_F T_F^2 \Nf^2 ~-~ 839808 C_F^2 T_F \Nf \right]
 \frac{a^4}{419904} ~+~ O(a^5)
\label{betamomggg}
\end{eqnarray}
\begin{eqnarray}
\beta^{\mbox{$\MOMqs$}}(a,0) &=& -~ \left[ \frac{11}{3} C_A - \frac{4}{3} 
T_F \Nf \right] a^2 ~-~ \left[ \frac{34}{3} C_A^2 - 4 C_F T_F \Nf 
- \frac{20}{3} C_A T_F \Nf \right] a^3 \nonumber \\
&& +~ \left[ \left[ 
203148 \pi \sqrt{3} \ln(3)
- 16929 \pi \sqrt{3} \ln^2(3) 
+ 18183 \pi^3 \sqrt{3} 
+ 214434 (\psi^\prime(\third))^2
\right. \right. \nonumber \\
&& \left. \left. ~~~~~
-~ 285912 \pi^2 \psi^\prime(\third) 
- 120798 \psi^\prime(\third)
- 23463 \psi^{\prime\prime\prime}(\third) 
- 7313328 s_2(\pisix)
\right. \right. \nonumber \\
&& \left. \left. ~~~~~
+~ 14626656 s_2(\pitwo)
+ 12188880 s_3(\pisix)
- 9751104 s_3(\pitwo)
+ 157872 \pi^4 
\right. \right. \nonumber \\
&& \left. \left. ~~~~~
+~ 80532 \pi^2 
+ 598752 \Sigma 
+ 1812294 \zeta(3) 
- 10781910
\right] C_A^3 \right. \nonumber \\
&& \left. ~~~~~+ \left[ 
49896 \pi \sqrt{3} \ln^2(3) 
- 598752 \pi \sqrt{3} \ln(3)
- 53592 \pi^3 \sqrt{3} 
- 465696 (\psi^\prime(\third))^2
\right. \right. \nonumber \\
&& \left. \left. ~~~~~~~~~
+~ 620928 \pi^2 \psi^\prime(\third) 
+ 5855328 \psi^\prime(\third)
- 14256 \psi^{\prime\prime\prime}(\third) 
+ 21555072 s_2(\pisix)
\right. \right. \nonumber \\
&& \left. \left. ~~~~~~~~~
-~ 43110144 s_2(\pitwo)
- 35925120 s_3(\pisix)
+ 28740096 s_3(\pitwo)
- 168960 \pi^4 
\right. \right. \nonumber \\
&& \left. \left. ~~~~~~~~~
-~ 3903552 \pi^2 
- 513216 \zeta(3) 
- 478224
\right] C_A^2 C_F \right. \nonumber \\
&& \left. ~~~~~+ \left[ 
1404 \pi \sqrt{3} \ln^2(3) 
- 16848 \pi \sqrt{3} \ln(3)
- 1508 \pi^3 \sqrt{3} 
- 77976 (\psi^\prime(\third))^2
\right. \right. \nonumber \\
&& \left. \left. ~~~~~~~~~
+~ 103968 \pi^2 \psi^\prime(\third) 
- 1176984 \psi^\prime(\third)
+ 18036 \psi^{\prime\prime\prime}(\third) 
+ 606528 s_2(\pisix)
\right. \right. \nonumber \\
&& \left. \left. ~~~~~~~~~
-~ 1213056 s_2(\pitwo)
- 1010880 s_3(\pisix)
+ 808704 s_3(\pitwo)
- 82752 \pi^4 
\right. \right. \nonumber \\
&& \left. \left. ~~~~~~~~~
+~ 784656 \pi^2 
- 217728 \Sigma
+ 1735992 \zeta(3) 
+ 9399240
\right] C_A^2 T_F \Nf \right. \nonumber \\
&& \left. ~~~~~+ \left[ 
9504 \pi \sqrt{3} \ln^2(3) 
- 114048 \pi \sqrt{3} \ln(3)
- 10208 \pi^3 \sqrt{3} 
+ 266112 (\psi^\prime(\third))^2
\right. \right. \nonumber \\
&& \left. \left. ~~~~~~~~~
-~ 354816 \pi^2 \psi^\prime(\third) 
- 4162752 \psi^\prime(\third)
+ 85536 \psi^{\prime\prime\prime}(\third) 
+ 4105728 s_2(\pisix)
\right. \right. \nonumber \\
&& \left. \left. ~~~~~~~~~
-~ 8211456 s_2(\pitwo)
- 6842880 s_3(\pisix)
+ 5474304 s_3(\pitwo)
- 109824 \pi^4 
\right. \right. \nonumber \\
&& \left. \left. ~~~~~~~~~
+~ 2775168 \pi^2 
- 342144 \Sigma
- 4790016 \zeta(3) 
+ 1283040
\right] C_A C_F^2 \right. \nonumber \\
&& \left. ~~~~~+ \left[ 
217728 \pi \sqrt{3} \ln(3)
- 18144 \pi \sqrt{3} \ln^2(3) 
+ 19488 \pi^3 \sqrt{3} 
+ 169344 (\psi^\prime(\third))^2
\right. \right. \nonumber \\
&& \left. \left. ~~~~~~~~~
-~ 225792 \pi^2 \psi^\prime(\third) 
- 1778112 \psi^\prime(\third)
+ 5184 \psi^{\prime\prime\prime}(\third) 
- 7838208 s_2(\pisix)
\right. \right. \nonumber \\
&& \left. \left. ~~~~~~~~~
+~ 15676416 s_2(\pitwo)
+ 13063680 s_3(\pisix)
- 10450944 s_3(\pitwo)
+ 61440 \pi^4 
\right. \right. \nonumber \\
&& \left. \left. ~~~~~~~~~
+~ 1185408 \pi^2 
- 3919104 \zeta(3) 
+ 3584736
\right] C_A C_F T_F \Nf \right. \nonumber \\
&& \left. ~~~~~+ \left[ 
1728 \pi \sqrt{3} \ln^2(3) 
- 20736 \pi \sqrt{3} \ln(3)
- 1856 \pi^3 \sqrt{3} 
\right. \right. \nonumber \\
&& \left. \left. ~~~~~~~~~
+~ 490752 \psi^\prime(\third)
- 3456 \psi^{\prime\prime\prime}(\third) 
+ 746496 s_2(\pisix)
\right. \right. \nonumber \\
&& \left. \left. ~~~~~~~~~
-~ 1492992 s_2(\pitwo)
- 1244160 s_3(\pisix)
+ 995328 s_3(\pitwo)
\right. \right. \nonumber \\
&& \left. \left. ~~~~~~~~~
+~ 9216 \pi^4 
- 327168 \pi^2 
- 870912 \zeta(3) 
- 1757376
\right] C_A T_F^2 \Nf^2 \right. \nonumber \\
&& \left. ~~~~~+ \left[ 
41472 \pi \sqrt{3} \ln(3)
- 3456 \pi \sqrt{3} \ln^2(3) 
+ 3712 \pi^3 \sqrt{3} 
- 96768 (\psi^\prime(\third))^2
\right. \right. \nonumber \\
&& \left. \left. ~~~~~~~~~
+~ 129024 \pi^2 \psi^\prime(\third)
+ 1389312 \psi^\prime(\third)
- 31104 \psi^{\prime\prime\prime}(\third) 
- 1492992 s_2(\pisix)
\right. \right. \nonumber \\
&& \left. \left. ~~~~~~~~~
+~ 2985984 s_2(\pitwo)
+ 2488320 s_3(\pisix)
- 1990656 s_3(\pitwo)
+ 39936 \pi^4 
\right. \right. \nonumber \\
&& \left. \left. ~~~~~~~~~
-~ 926208 \pi^2 
+ 124416 \Sigma
+ 1741824 \zeta(3) 
+ 513216
\right] C_F^2 T_F \Nf \right. \nonumber \\
&& \left. ~~~~~+ \left[ 
55296 \pi^2
- 82944 \psi^\prime(\third)
+ 1492992 \zeta(3)
- 1057536
\right] C_F T_F^2  \Nf^2 \Nf \right] \frac{a^4}{69984} \nonumber \\
&& +~ O(a^5)
\label{betamomq}
\end{eqnarray}
and
\begin{eqnarray}
\beta^{\mbox{$\MOMhs$}}(a,0) &=& -~ \left[ \frac{11}{3} C_A - \frac{4}{3} 
T_F \Nf \right] a^2 ~-~ \left[ \frac{34}{3} C_A^2 - 4 C_F T_F \Nf 
- \frac{20}{3} C_A T_F \Nf \right] a^3 \nonumber \\
&& +~ \left[ \left[ 
97416 \pi \sqrt{3} \ln^2(3) 
- 1168992 \pi \sqrt{3} \ln(3)
- 104632 \pi^3 \sqrt{3} 
+ 14850 (\psi^\prime(\third))^2
\right. \right. \nonumber \\
&& \left. \left. ~~~~~
-~ 19800 \pi^2 \psi^\prime(\third) 
+ 7112448 \psi^\prime(\third)
+ 35343 \psi^{\prime\prime\prime}(\third) 
+ 42083712 s_2(\pisix)
\right. \right. \nonumber \\
&& \left. \left. ~~~~~
-~ 84167424 s_2(\pitwo)
- 70139520 s_3(\pisix)
+ 56111616 s_3(\pitwo)
- 87648 \pi^4 
\right. \right. \nonumber \\
&& \left. \left. ~~~~~
-~ 4741632 \pi^2 
- 85536 \Sigma 
- 1689336 \zeta(3) 
- 35200008
\right] C_A^3 \right. \nonumber \\
&& \left. ~~~~~+~ \left[ 
881280 \pi \sqrt{3} \ln(3)
- 73440 \pi \sqrt{3} \ln^2(3) 
+ 78880 \pi^3 \sqrt{3} 
- 5400 (\psi^\prime(\third))^2
\right. \right. \nonumber \\
&& \left. \left. ~~~~~~~~~~
+~ 7200 \pi^2 \psi^\prime(\third)
- 5593536 \psi^\prime(\third)
- 12852 \psi^{\prime\prime\prime}(\third) 
- 31726080 s_2(\pisix)
\right. \right. \nonumber \\
&& \left. \left. ~~~~~~~~~~
+~ 63452160 s_2(\pitwo)
+ 52876800 s_3(\pisix)
- 42301440 s_3(\pitwo)
+ 31872 \pi^4 
\right. \right. \nonumber \\
&& \left. \left. ~~~~~~~~~~
+~ 3729024 \pi^2 
+ 31104 \Sigma 
+ 11562912 \zeta(3) 
+ 29167776
\right] C_A^2 T_F \Nf \right. \nonumber \\
&& \left. ~~~~~+~ \left[ 
13824 \pi \sqrt{3} \ln^2(3) 
- 165888 \pi \sqrt{3} \ln(3)
- 14848 \pi^3 \sqrt{3} 
\right. \right. \nonumber \\
&& \left. \left. ~~~~~~~~~~
+~ 1057536 \psi^\prime(\third)
+ 5971968 s_2(\pisix)
- 11943936 s_2(\pitwo)
- 9953280 s_3(\pisix)
\right. \right. \nonumber \\
&& \left. \left. ~~~~~~~~~~
+ 7962624 s_3(\pitwo)
- 705024 \pi^2 
- 3981312 \zeta(3) 
- 4105728
\right] C_A T_F^2 \Nf^2 \right. \nonumber \\
&& \left. ~~~~~+~ \left[ 
103680 \pi^2
- 155520 \psi^\prime(\third)
- 16422912 \zeta(3)
+ 18817920
\right] C_A C_F T_F \Nf \right. \nonumber \\
&& \left. ~~~~~+~ \left[ 
5971968 \zeta(3)
- 5723136
\right] C_F T_F^2 \Nf^2 - 559872 C_F^2 T_F \Nf \right] \frac{a^4}{279936}
\nonumber \\
&& +~ O(a^5)
\label{betamomh}
\end{eqnarray}
where $\psi(z)$ is the derivative of the logarithm of the Euler 
$\Gamma$-function, $\Nf$ is the number of massless quarks and $C_F$, $C_A$ and 
$T_F$ are the usual colour group Casimirs. Aside from the usual class of 
numbers which  appear in the renormalization group functions such as rationals 
and the Riemann zeta function, $\zeta(z)$, numbers deriving from the basic one 
and two loop scalar master diagrams computed in \cite{11,12,13,14} also occur 
which are 
\begin{equation}
s_n(z) ~=~ \frac{1}{\sqrt{3}} \Im \left[ \mbox{Li}_n \left(
\frac{e^{iz}}{\sqrt{3}} \right) \right] ~~~,~~~ 
\Sigma ~=~ {\cal H}^{(2)}_{31} ~+~ {\cal H}^{(2)}_{43}
\end{equation}
where $\mbox{Li}_n(z)$ is the polylogarithm function and $\Sigma$ is a 
particular combination of harmonic polylogarithms, \cite{14,28}.

Having computed the three loop $\beta$-functions it is worthwhile evaluating 
them numerically in order to compare with \cite{10}. We find 
\begin{eqnarray}
\beta^{\mbox{$\MOMgggs$}}(a,0) &=& -~ ( 11.0000000 - 0.6666667 \Nf ) a^2 ~-~
( 102.0000000 - 12.6666667 \Nf ) a^3 \nonumber \\
&& -~ ( 1570.9843804 + 0.5659290 \Nf - 67.0895364 \Nf^2 + 2.6581155 \Nf^3 ) 
a^4 \nonumber \\
&& +~ O(a^5) \nonumber \\
\beta^{\mbox{$\MOMqs$}}(a,0) &=& -~ ( 11.0000000 - 0.6666667 \Nf ) a^2 ~-~
( 102.0000000 - 12.6666667 \Nf ) a^3 \nonumber \\
&& -~ ( 1843.6527285 - 588.6548455 \Nf + 22.5878118 \Nf^2 ) a^4 ~+~ O(a^5)  
\nonumber \\
\beta^{\mbox{$\MOMhs$}}(a,0) &=& -~ ( 11.0000000 - 0.6666667 \Nf ) a^2 ~-~
( 102.0000000 - 12.6666667 \Nf ) a^3 \nonumber \\
&& -~ ( 2813.4929484 - 617.6471542 \Nf + 21.5028181 \Nf^2 ) a^4 ~+~ O(a^5)  
\end{eqnarray}
for $SU(3)$. Clearly the expressions have the same structure from the point of 
view of the $\Nf$ dependence as \cite{10}. As the term in the polynomial in 
$\Nf$ with the largest error estimate in \cite{10} was the $\Nf$ independent 
term, then we can use that term for comparison. Therefore, taking the ratio of 
the central value of the estimates in \cite{10} to those of the exact 
expressions, (\ref{betamomggg}), (\ref{betamomq}) and (\ref{betamomh}), we find
that the percentage errors are $2.227$\%, $0.184$\% and $1.954$\% for MOMggg, 
MOMq and MOMh respectively. Clearly MOMq is the most accurate whilst MOMggg was
the worst. Though it is worth noting that whilst the $\Nf$ independent 
coefficient of the MOMggg three loop term had a relatively large $8$\% error 
estimate the exact result was comfortably within this. Also there was a large 
uncertainty for the $\Nf$ coefficient in this scheme but we find the same sign 
where the actual coefficient turns out to be very small. Indeed this probably 
reflects the accidental cancellation noted in \cite{10} for this term. We 
obtain the precise coefficient for the cubic term in $\Nf$ in the MOMggg case. 
Despite the larger discrepancy for this scheme it is still a testimony to the 
ingenuity of the authors of \cite{10} and the software and hardware technology 
of over a decade ago that results were produced which were very close to the 
exact answer for all three schemes. As \cite{10} also commented on a lattice 
based operator product expansion prediction, \cite{29}, of our benchmark 
coefficient for the MOMggg scheme in relation to the corresponding 
$\widetilde{\mbox{MOM}}$gg scheme coefficient, we note that for comparison we
have
\begin{equation}
\left. \frac{}{} \beta_2^{\mbox{$\MOMgggs$}} \right|_{\Nf = 0} ~=~
0.6512787 \left. \frac{}{} \beta_2^{\mbox{$\wMOMggs$}} \right|_{\Nf = 0} 
\end{equation}
which is very close and well within the error of the estimate of $0.64(5)$ of 
\cite{10} than the $1.5(3)$ of \cite{29}.  

We conclude with brief remarks. We have provided the explicit forms of the QCD
$\beta$-function in the MOMggg, MOMq and MOMh renormalization schemes in the
Landau gauge where those for an arbitrary linear covariant gauge will be 
provided in \cite{23}. The results for all three schemes are consistent to
within a few percent of the numerical estimates of \cite{10} which involved a 
resummation based on the $2$-point function substructure at the symmetric 
subtraction point. Given this the method of \cite{10} could actually prove 
useful now for determining a numerical estimate of the {\em four} loop MOM 
scheme $\beta$-functions. This is partly due to the fact that \cite{10} used 
{\sc Mincer}, \cite{15}, which can handle the three loop computations required 
to extend the coupling constant mapping to the next order, but also because 
computer power has increased significantly since the appearance of \cite{10}. 
Moreover, with analytic results now available for the three MOM schemes, there 
is independent information which will provide intermediate checks on such 
numerical computations.


\begin{thebibliography}{99}
\bibitem{1} G. 't Hooft, Nucl. Phys. {\bf B61} (1973), 455.
\bibitem{2} W.A. Bardeen, A.J. Buras, D.W. Duke \& T. Muta, Phys. Rev.
{\bf D18} (1978), 3998.
\bibitem{3} T. van Ritbergen, J.A.M. Vermaseren \& S.A. Larin, Phys. Lett.
{\bf B400} (1997), 379.
\bibitem{4} M. Czakon, Nucl. Phys. {\bf B710} (2005), 485.
\bibitem{5} K.G. Chetyrkin, A.L. Kataev \& F.V. Tkachov, Nucl. Phys. {\bf
B174} (1980), 345.
\bibitem{6} K.G. Chetyrkin \& F.V. Tkachov, Nucl. Phys. {\bf B192} (1981), 159.
\bibitem{7} A.N. Vasil'ev, Yu.M. Pis'mak \& Yu.R. Honkonen, Theor. Math. Phys.
{\bf 47} (1981), 465.
\bibitem{8} J.A.M. Vermaseren, math-ph/0010025. 
\bibitem{9} W. Celmaster \& R.J. Gonsalves, Phys. Rev. {\bf D20} (1979), 1420.
\bibitem{10} K.G. Chetyrkin \& T. Seidensticker, Phys. Lett. {\bf B495} (2000),
74. 
\bibitem{11} A.I. Davydychev, J. Phys. {\bf A25} (1992), 5587.
\bibitem{12} N.I. Usyukina \& A.I. Davydychev, Phys. Atom. Nucl. {\bf 56}
(1993), 1553.
\bibitem{13} N.I. Usyukina \& A.I. Davydychev, Phys. Lett. {\bf B332} (1994),
159.
\bibitem{14} T.G. Birthwright, E.W.N. Glover \& P. Marquard, JHEP {\bf 0409}
(2004), 042.
\bibitem{15} S.G. Gorishny, S.A. Larin, L.R. Surguladze \& F.K. Tkachov,
Comput. Phys. Commun. {\bf 55} (1989), 381.
\bibitem{16} D.J. Gross \& F.J. Wilczek, Phys. Rev. Lett. {\bf 30}
(1973), 1343.
\bibitem{17} H.D. Politzer, Phys. Rev. Lett. {\bf 30} (1973), 1346.
\bibitem{18} D.R.T. Jones, Nucl. Phys. {\bf B75} (1974), 531.
\bibitem{19} W.E. Caswell, Phys. Rev. Lett. {\bf 33} (1974), 244.
\bibitem{20} O.V. Tarasov, A.A. Vladimirov \& A.Yu. Zharkov, Phys. Lett.
{\bf B93} (1980), 429.
\bibitem{21} S.A. Larin \& J.A.M. Vermaseren, Phys. Lett. {\bf B303} (1993),
334.
\bibitem{22} S. Laporta, Int. J. Mod. Phys. {\bf A15} (2000), 5087.
\bibitem{23} J.A. Gracey, paper in preparation. 
\bibitem{24} A.I. Davydychev \& P. Osland, Phys. Rev. {\bf D59} (1999), 014006.
\bibitem{25} C. Studerus, Comput. Phys. Commun. {\bf 181} (2010), 1293.
\bibitem{26} C.W. Bauer, A. Frink \& R. Kreckel, cs/0004015.
\bibitem{27} P. Nogueira, J. Comput. Phys. {\bf 105} (1993), 279. 
\bibitem{28} L.G. Almeida \& C. Sturm, Phys. Rev. {\bf D82} (2010), 054017.
\bibitem{29} P. Boucaud, A. Le Yaounac, J.P. Leroy, J. Micheli, O. Pene \&
J. Rodriguez-Quintero, Phys. Lett. {\bf B493} (2000), 315.
\end{thebibliography}
\end{document}